\documentclass[twocolumn,prl,showpacs,preprintnumbers,amsmath,amssymb,superscriptaddress]{revtex4}
\usepackage{graphicx}% Include figure files
\usepackage{dcolumn}% Align table columns on decimal point
\usepackage{bm}% bold math
\usepackage{color}
\font\scripti=cmmi7
\font\scriptscripti=cmmi5
\def\sib#1{\setbox0 = \hbox{\scripti #1}
  \kern-.02em\copy0\kern-\wd0
  \kern.04em\box0} % script italic bold 
\def\ssib#1{\setbox0 = \hbox{\scriptscripti #1}
  \kern-.02em\copy0\kern-\wd0
  \kern.04em\box0} % scriptscript italic bold
\font\tenib=cmmib10 % italic bold for math
\skewchar\tenib='177 \skewchar\tenib='177 \skewchar\tenib='177
\def\pbold#1{\setbox0 = \hbox{$ #1 $}
  \kern-.022em\copy0\kern-\wd0
  \kern.011em\copy0\kern-\wd0
  \kern.011em\copy0\kern-\wd0
  \kern.011em\copy0\kern-\wd0
  \kern.011em\box0} % poorman's bold
\usepackage{graphicx}% Include figure files
\usepackage{dcolumn}% Align table columns on decimal point
\usepackage{bm}% bold math

\def\up{\uparrow}
\def\dwn{\downarrow}

\def\lesssim{\ \raise.3ex\hbox{$<$}\kern-0.8em\lower.7ex\hbox{$\sim$}\ }
\def\gesim{\ \raise.3ex\hbox{$>$}\kern-0.8em\lower.7ex\hbox{$\sim$}\ }

\begin{document}
\title{Collisional Dynamics of Polaronic Clouds Immersed in a Fermi Sea}
\author{Hiroyuki Tajima}
\affiliation{Department of Mathematics and Physics, Kochi University, Kochi 780-8520, Japan}
\affiliation{Quantum Hadron Physics Laboratory, RIKEN Nishina Center, Wako, Saitama 351-0198, Japan}

\author{Junichi Takahashi}
\affiliation{Departement of Electronic and Physical Systems, Waseda University, Tokyo 169-8555, Japan}

\author{Eiji Nakano}
\affiliation{Department of Mathematics and Physics, Kochi University, Kochi 780-8520, Japan}
\affiliation{Institut f\"{u}r Theoretische Physik, Goethe Universit\"{a}t Frankfurt, D-60438 Frankfurt am Main, Germany}

\author{Kei Iida}
\affiliation{Department of Mathematics and Physics, Kochi University, Kochi 780-8520, Japan}

\date{\today}
\begin{abstract}
We propose a new protocol to examine many-polaron properties in a cold atom experiment.
Initially, polaronic clouds are prepared around the opposite edges of a majority gas cloud.
After time evolution, the collision of two clouds exhibits various polaronic effects.
To see how {\it collective} properties of many polarons with mediated interactions appear in the case in which the impurity and majority gases are composed of mass-balanced fermions with different spin components, we perform a nonlinear hydrodynamic simulation for collisional dynamics of two Fermi polaronic clouds.
We found that the dynamics is governed by the impurity Fermi pressure, polaron energy, and multi-polaron correlations.  
In particular, shock waves occur in such a way as to reflect the many-body properties of polarons through the first sound of minority clouds. 
%The different density dependence among these effects would help to confirm each effect by experiments. 
Our idea is applicable to other systems such as Bose polarons as well as mass-imbalanced mixtures.
\end{abstract}
%\pacs{03.75.Ss, 03.75.-b, 03.70.+k}
\maketitle
%%%%%%%%%%%%%%%%%%%%%%%%%%%%%%%%%%%%%%%%%%%%%%%%%%%%%%%%%%%%%%%%%%%%%%%%%%%%%
%\par
%\section{Introduction}

%\section{Collision of impurity clouds in a Fermi sea}
{\it Introduction}---
Ultracold atomic gases provide us with an ideal testing ground for fundamental quantum physics~\cite{Giorgini,Bloch}.
Recently, this system enables quantum simulation of extreme states of matter such as neutron matter~\cite{Horikoshi1,Navon,Nascimbene,Ku,Horikoshi2,Horikoshi3,Heiselberg,Gezerlis,Tajima,Pieter,Strinati,OhashiR} and high-$T_{\rm c}$ superconductors~\cite{Hofstetter,Chin,Zoller,Mazurenko,Gross,Koepsell}. 
A striking feature of this system is the controllability of physical parameters such as interaction and density.
This advantage helps to extensively investigate various properties of
the most fundamental quantum impurity system, namely, a polaron~\cite{Massignan,Chevy}.
\par
The Fermi (Bose) polaron can be realized by immersing impurity atoms into fermionic (bosonic) gas clouds, which work as a medium for impurity atoms. 
Since these systems are described by a simple Hamiltonian,
various physical properties of the polarons can be
investigated experimentally and theoretically as a benchmark of quantum many-body physics.
In particular, the excitation properties of polarons have been precisely determined from radio-frequency (RF) spectroscopies~\cite{MIT1,Kohstall,JILA,Aarhus,LENS2,Boiling} and the spin-dipole oscillation~\cite{ENS2,MIT2}.
Furthermore, the induced polaron-polaron interaction in a Fermi sea has been observed in recent experiments~\cite{CChin,Edri}.
This interaction, which reflects medium properties, has attracted much attention from various research fields such as condensed matter~\cite{Brando} and nuclear physics~\cite{Schwenk,Pethick}.
For example, light cluster and strange hadronic states in nuclear matter can be regarded as quantum impurities in strongly correlated nuclear media~\cite{Oertel}.
\par
The energy shift and broadening in Fermi polaron spectra have been discussed in connection with the polaron-polaron interaction~\cite{Chevy,Giraud,Mistakidis,TajimaUchino}.
In the case of Bose polarons, formation of bipolarons originating from these mediated interactions has been predicted~\cite{Naidon,Camacho,Zinner1,Zinner2,Zinner3}. 
While these induced interactions are relatively long-ranged, such as Ruderman-Kittel-Kasuya-Yoshida interaction~\cite{RK,Kasuya,Yosida} in heavy Fermi polarons~\cite{Nishida,Suchet} and Yukawa or Efimov attractions~\cite{Naidon,NaidonEndo} in Bose polarons,
such a non-locality of the induced interactions is still under investigation.
Although non-trivial pairing states due to the induced interactions have extensively been discussed~\cite{Nishida,Bulgac,Kalas,Sheehy,Kinnunen}, these states have not yet been observed.
The non-locality may also play a crucial role in the spatial structure of impurities and media~\cite{Aalto,Tylutki,Takahashi2,Watanabe,Nakano}. 
In examining how induced interactions depend on the inter-impurity distance, an alternative experiment design associated with the {multi-polaron} dynamics, {which could directly reflect the influence of the induced interactions as compared with the static properties}, is promising~\cite{Mistakidis2,Mistakidis3,Mistakidis4,Hofmann,Charalambous}.
%Indeed, the observation of the collective dynamics enable us to investigate strong-coupling effects in ultracold Fermi gases~\cite{}.

\begin{figure}[t]
\begin{center}
\includegraphics[width=7cm]{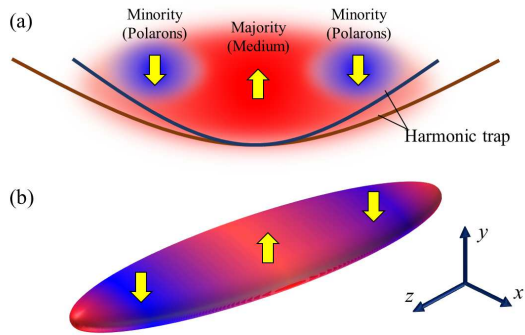}
\end{center}
\caption{ (a) Initial setup for later polaronic cloud collision in our protocol.
As schematically shown in (b) in a coordinate system, two polaronic minority ($\dwn$) clouds are separately prepared near the opposite edges of a medium majority gas cloud ($\up$) in an elongated harmonic trap.
%After certain time evolution, these impurity clouds fall into the trap center due to the harmonic trap potential as well as the polaron binding energy.
}
\label{fig1}
\end{figure}
In this work, we propose a scheme to investigate collective properties of polarons in cold atom experiments, as shown in Fig. \ref{fig1}. 
First, we prepare two-component gas clouds in such a way that two polaronic clouds of minority component (spin down) embedded in a majority gas cloud are separated at a distance in the axial direction.
Since the minority clouds gather at the trap center after certain time evolution, they collide with each other.
Since the spin-selective preparation in real space is realizable in ultracold atoms {such as $^6$Li mixtures with two hyperfine states}~\cite{Thomas,MIT3,LENS4},
our idea can be demonstrated in future experiments.
{In particular, a spatially selective RF pulse~\cite{Sagi} for the transition from {the minority to the majority spin} near the trap center{, as well as} the spatial control of interspecies interaction~\cite{Arunkumar} can be applied for the intial setup}.
We note that a similar protocol has theoretically been proposed to realize a spin-polarized droplet in a unitary superfluid Fermi gas~\cite{Magierski}. 
\par
{\it Formulation---}To see non-equilibrium collective properties of polarons in our setup, we demonstrate the collisional dynamics of polaronic clouds by solving non-linear hydrodynamic equations and discuss how the polaronic properties affect the dynamics.
At first, we do not specify any parameters such as {minority-majority} interaction since our approach can generally be applied to various setups as long as the system is in the hydrodynamic regime {where the relaxation time is shorter than the inverse of the mode frequency~\cite{PinesNozieres}}.
After overviewing our framework, we focus on mass-balanced unitary Fermi polarons as a typical example.
We note that the hydrodynamic description works well to reproduce the dynamics of strongly interacting Fermi gases in experiments~\cite{Thomas,ENS2,MIT2,Cao,Baird}.
Euler's equation relevant for such collisional dynamics is given by~\cite{MIT3,Taylor,Chiacchiera}
\begin{eqnarray}
\label{eq1}
\frac{\partial \bm{v}_\sigma}{\partial t}+\frac{\nabla v_\sigma^2}{2}=-\frac{\nabla}{m_{\sigma}^*}\left(\frac{\partial E}{\partial n_\sigma}+V_{\rm trap,\sigma}\right)-\gamma_\sigma (\bm{v}_\sigma-\bm{v}_{-\sigma}),
\end{eqnarray}
\begin{eqnarray}
\label{eq2}
\frac{\partial n_{\sigma}}{\partial t}+\nabla\cdot(n_{\sigma} \bm{v}_\sigma)=0,
\end{eqnarray}
where $n_\sigma$ and $\bm{v}_\sigma$ are the local density and velocity field of majority ($\sigma=\up$) and minority ($\sigma=\dwn$) atomic clouds, $V_{{\rm trap},\sigma}(z,r_{\perp})=(m_\sigma\omega_{z,\sigma}^2 z^2 + m_\sigma\omega_{\perp,\sigma}^2 r_\perp^2)/2$ with $r_\perp=\sqrt{x^2+y^2}$ is the trap potential for $\sigma$ component of bare mass $m_\sigma$ located at the position $(x,y,z)$ with respect to the trap center, and
$\gamma_\sigma$ and $m_\sigma^*$ are the spin relaxation coefficient and effective mass for $\sigma$ component~{\cite{note1}}.
For the energy density $E$, we employ the ground-state form based on the Landau-Pomeranchuk Hamiltonian~\cite{Pilati} as
\begin{eqnarray}
\label{eq3}
E=\Xi_{\rm P}+\Xi_{\rm A}+\Xi_{\rm F}+\Xi_{\rm G}+O(n_\dwn^4),
\end{eqnarray}
where
$\Xi_{\rm P}$ is the degeneracy pressure term. 
%(hereafter we denote ``P"). 
In the case of Fermi-Fermi mixtures, it is given by  
%\begin{eqnarray}
%\label{eq4}
$\Xi_{\rm P}=\frac{3}{5}n_{\up}\varepsilon_{\rm F,\up}+\frac{3}{5}n_{\dwn}\varepsilon_{\rm F,\dwn},$
%\end{eqnarray}
where $\varepsilon_{{\rm F},\sigma}=(6\pi^2n_{\sigma})^{\frac{2}{3}}/(2m_\sigma^*)$.
%\red{We note that $\Xi_{\rm P}$ would be replaced by boson-boson repulsion in the case of bosonic medium.}
The majority component can approximately be described by an ideal gas; we thus set
%the majority effective mass is equal to the bare atomic mass, {\bf and also} the relaxation rate is negligibly small, {\bf i.e.,}
$m_{\up}^*=m_\up$.
We note that this majority cloud is incompressible enough for the gradient term in the right hand side of Eq.~(\ref{eq1}) for $\sigma=\up$ to dominate.
In this sense, we set $\bm{v}_{\up}=\bm{0}$ and $\gamma_\up=0$ and the majority density is approximated by the Thomas-Fermi density profile of an ideal trapped Fermi gas as
$n_{\up}=\frac{(2m_\up)^{\frac{3}{2}}}{6\pi^2}\left[E_{\rm F,0}-V_{\rm trap}(z,r_\perp)\right]^{\frac{3}{2}}$,
where $E_{\rm F,0}=(6N_\up\omega_z\omega_\perp^2)^{1/3}$ is the Fermi energy of majority atoms of total number $N_{\up}$.
The other terms 
%\begin{eqnarray}
%\label{eq5}
$\Xi_{\rm A}=-\alpha n_{\dwn}$,
%\end{eqnarray}
%\begin{eqnarray}
%\label{eq6}
$\Xi_{\rm F}=\zeta n_{\dwn}^2$,
%\end{eqnarray}
%\begin{eqnarray}
%\label{eq6-2}
and $\Xi_{\rm G}=\tau n_{\dwn}^3$
%\end{eqnarray}
are associated with the polaron energy, polaron-polaron interaction, and induced three-body force, respectively.
{We note that $\Xi_{\rm A}$ also corresponds to the peak position in the RF spectroscopy in the single-impurity limit.} 
From dimensional analysis, one can obtain $\alpha=\chi\varepsilon_{\rm F,\up}$,
$\zeta=\frac{3}{5}\kappa\frac{\varepsilon_{\rm F,\up}}{n_{\up}}$, and $\tau=\frac{3}{5}\lambda\frac{\varepsilon_{\rm F,\up}}{n_{\up}^2}$ with constants $\chi$, $\kappa$, and $\lambda$. 
{{Here we take it for granted} that these constants originate from strong {(unitary limit) coupling} between two spins and medium {polarization}. However, a precise characterization of them is out of scope in this paper.
We employ these values obtained in the previous work as mentioned below.}
\par
Before moving to numerical calculations,
we clarify how the polaronic properties appear in the collisional dynamics.
$\Xi_{\rm P}$ induces the repulsive force as
\begin{eqnarray}
\label{eq7}
\nabla\left(\frac{\partial \Xi_{\rm P}}{\partial n_\dwn}\right)= \frac{(6\pi^2)^{\frac{2}{3}}}{3m_\dwn^*}\frac{\nabla n_\dwn}{n_\dwn^{\frac{1}{3}}},
\end{eqnarray}
which depends only on the impurity density $n_\dwn$.
On the other hand, an attractive force originating from $\Xi_{\rm A}$ as given by
\begin{eqnarray}
\label{eq8}
\nabla\left(\frac{\partial \Xi_{\rm A}}{\partial n_\dwn}\right)=-\frac{(6\pi^2)^{\frac{2}{3}}\chi}{3m_\up}\frac{\nabla n_\up}{n_\up^{\frac{1}{3}}}
\end{eqnarray}
depends only on the majority density $n_\up$.
Therefore, $\Xi_{\rm A}$ works as a potential energy for impurities.
Moreover, the interaction terms $\Xi_{\rm F}$ and $\Xi_{\rm G}$ depend on both of the minority and majority densities as
\begin{eqnarray}
\label{eq9}
\nabla\left(\frac{\partial \Xi_{\rm F}}{\partial n_\dwn}\right)=\frac{(6\pi^2)^{\frac{2}{3}}\kappa}{5m_\up n_{\up}^{\frac{1}{3}}}\left[3\nabla n_\dwn - \frac{n_\dwn}{n_\up}\nabla n_\up\right],
\end{eqnarray}
\begin{eqnarray}
\label{eq10}
\nabla\left(\frac{\partial \Xi_{\rm G}}{\partial n_\dwn}\right)=
\frac{(6\pi^2)^{\frac{2}{3}}\lambda}{5m_\up n_{\dwn}^{\frac{1}{3}}}\left[9\frac{n_\dwn}{n_\up}\nabla n_\dwn-2\left(\frac{n_\dwn}{n_\up}\right)^2\nabla n_\up\right].
\end{eqnarray}
From the above, one can find that effects of the induced multi-polaron forces (\ref{eq9}) and (\ref{eq10}) on the dynamics become stronger in a region where the gradient of the majority density profile $|\nabla n_\up|$ is larger.
We emphasize that it is a sharp contrast to the case of $\Xi_{\rm P}$ in which the force is not affected by the majority density.
Such different density dependences enable us to selectively examine each effect during the dynamics in an appropriate setup.
This is remarkable since recently a variety of shapes of trap potentials are experimentally realized~{\cite{Mukherjee}}.
If one prepares the majority density profile with a large gradient and locates impurity clouds on such a gradient,
only polaronic effects on the driving force would be strongly enhanced. 
\par
From now on, we consider mass-balanced Fermi polarons ($m_\up=m_\dwn$) with the $\up$-$\dwn$ attraction at unitarity.  As long as the trap potential is independent of $\sigma$, i.e., $V_{{\rm trap},\dwn}=V_{{\rm trap},\up}\equiv V_{\rm trap}$ with $\omega_{z,\dwn}=\omega_{z,\up}\equiv\omega_z$ and $\omega_{\perp,\dwn}=\omega_{\perp,\up}\equiv\omega_\perp$, quasi-particle properties of this system are well-known theoretically~\cite{Pilati,Chevy2,Combescot,Schmidt:2011zu,PhysRevA.95.013612,Prokofev20081,Prokofev20082,Vlietinck2013,Kroiss2015,PhysRevA.94.051605,Bruun,TajimaUchino2,Mulkerin} and experimentally~\cite{MIT1,LENS2,Boiling,ENS2,MIT2}.
%Taking into account these previous works, 
We thus employ $\chi=0.6$, $\kappa=0.2$, and $m_\dwn^*/m_\dwn=1.17$.
Note that these quantities can be approximately deduced from a variational theory that involves one particle-hole excitation near the Fermi sea~\cite{Chevy}.
%It is a strong advantage compared to the previous work~\cite{Taylor}, which use the lowest order constraint variational (LOCV) approximation to describe the interaction term in the upper-branch dynamics.
%Particularly, these are close to the FNDMC results ($\chi=0.59$, $\kappa=0.14$, and $m_\dwn^*/m_\dwn=1.09$), which essentially involve effects of the polaron formation and the induced interpolaron correlations in a non-perturbative manner.
Although we take $\lambda=0$ since $\lambda$ is negligibly small in the present case~\cite{Pilati}, it plays a non-trivial role in Bose polarons~\cite{NaidonarXiv}.
%We note that the values of $\chi$ are also consistent with other theoretical studies~\cite{Schmidt:2011zu,PhysRevA.95.013612,Prokofev20081,Prokofev20082,Vlietinck2013,Kroiss2015,PhysRevA.94.051605}
%as well as the experiments~\cite{MIT1,LENS2,Boiling}.
The spin relaxation coefficient $\gamma_{\dwn}$ of a Fermi polaron is proportional to $T^2$ in the low temperature regime~\cite{Boiling,Bruun,TajimaUchino2,Mulkerin},
and thus ignored in this analysis.
%Although {\bf $\gamma_\dwn$} depends {\bf also} on the {\bf relative} velocity 
%{\bf $\bm{v}_\dwn-\bm{v}_\up$ in principle, it} can {\bf generally} be extracted in the {\bf experiments in a velocity-independent manner.}
%{\bf We can thus assume $\gamma_{\dwn}(r_\perp,z)=\rho\left[1-V_{\rm trap}(r_{\perp},z)/E_{\rm F,0}\right]^{-1}$}, where $\rho$ is a {\bf factor that behaves as} $T^2$. 
Note that the bulk viscosity is exactly zero at unitarity limit due to the conformal symmetry  {since the bulk viscosity characterizes the transport property when the fluid volume is expanded or compressed}~\cite{Fujii}. 
The shear viscosity, at least in an unpolarized unitary Fermi gas, is small due to strong correlations~\cite{Cao,Enss,Kagamihara}. 
It is expected to be close to the Kovton-Son-Starinets bound~\cite{KSS}.
%In the highly polarized case, it is deeply related to the polaron lifetime~\cite{Cai}, which is shortened by effects of the finite impurity concentration~\cite{LENS2,TajimaUchino}.
For simplicity, we ignore these viscosities as implied in Eq.~(\ref{eq1}).% in this work. 
\par
%To demonstrate our scenario, we numerically address non-equilibrium dynamics of separated impurity clouds.
To see a transient behavior before the collision, we focus on the axial motion (along which we take the $z$-direction).
The hydrodynamic equations relevant for this motion can be reduced to the one-dimensional  equations on the $z$ axis {(note however that the system itself {can be} regarded as {a bunch of locally three-dimensional subsystems})}.
Following recent polaron experiments~\cite{LENS2}, we take $\omega_z/\omega_\perp=20/233$ and $N_\up=1.5\times 10^5$.
We set an initial condition in such a way that the initial polaronic density profile $n_\dwn(z,t=0)\equiv n_{\dwn}(z,r_\perp,t)|_{r_\perp=t=0}$ has double peaks near the edge of the majority gas cloud of central density $n_{\up}(0)$ as in Fig.\ \ref{fig1}.
We adopt
\begin{eqnarray}
\label{eqInitial}
n_\dwn(z,t=0)=\frac{YR_zn_\up(0)}{2\eta\sqrt{2\pi}}\left[e^{-\frac{(z+z_{\rm I})^2}{2\eta^2}}+e^{-\frac{(z-z_{\rm I})^2}{2\eta^2}}\right],
\end{eqnarray} 
Here, we set $z_{\rm I}=0.8R_z$ unless otherwise noted.
Since the impurity concentration $Y$ is required to be small to keep the stability of the majority cloud, we take $Y=0.01$.
Finally, the width of impurity clouds is taken as $\eta =0.05R_z$.
%We have confirmed that change in this choice would make no qualitative difference.
The nonlinear differential equations (\ref{eq1}) and (\ref{eq2}) for polarons with $v_{x,y,\dwn}=\partial_{x,y}n_{\dwn}=0$ are numerically solved by the fourth-order Runge-Kutta method with the discretization $\Delta z=4\times10^{-5}R_z$ and $\Delta t=1.33\times 10^{-3}E_{\rm F,0}^{-1}$.   
These are chosen in such a way as to fulfill $\Delta z/\Delta t>|v_{z,\dwn}|$, which is required to stabilize the numerical simulation as indicated by the Courant-Friedrichs-Lewy condition~\cite{CFL}.
Furthermore, we take a smearing process as $f(z_i)=[f(z_{i-1})+f(z_{i+1})]/2$ ($f=n_{\dwn}, v_{\dwn}$) at each step.
%This procedure is justified when $\Delta z$ is sufficiently small. 
%We note that our results are qualitatively unchanged against the choice of these parameters.
\par

%\section{Preliminary numerical results of hydrodynamic collision of impurity clouds}
{\it Results}---
\begin{figure}[t]
\begin{center}
\includegraphics[width=0.9\hsize]{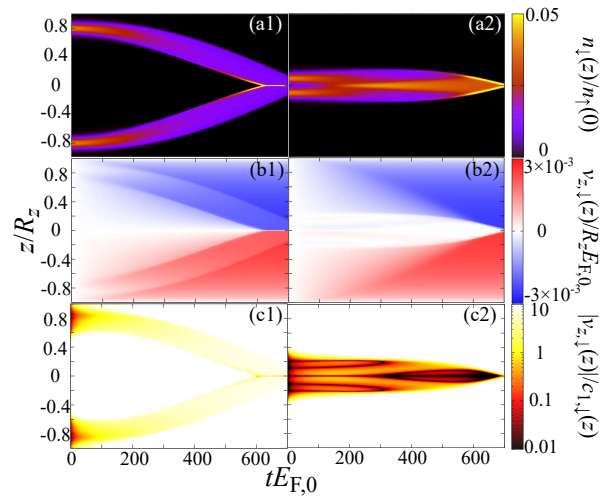}
\end{center}
\caption{(a1) Fermi polaron density profile $n_\dwn(z)$ and (b1) velocity field $v_{z,\dwn}(z)$ with $z_{\rm I}=0.8R_z$ in the presence of the impurity Fermi pressure ($\Xi_{\rm P}$) and polaronic effects ($\Xi_{\rm A}+\Xi_{\rm F}$).
{In the panel (c1), we show the Mach number profile $|v_{\dwn}(z)|/c_{1,\dwn}(z)$ of polaronic clouds. }
For comparison, we show the results with $z_{\rm I}=0.1R_z$ in the panels {(a2), (b2) and (c2). }
}
\label{fig2}
\end{figure}
%First, we demonstrate the collisional dynamics of polarized $^6$Li Fermi gases.
Figure~\ref{fig2} (a1) exhibits time-dependent axial density profile $n_{\dwn}(z)$ of Fermi polaronic clouds with $z_{\rm I}=0.8R_z$. 
The collision time $t_{\rm c}$ can be estimated as the $1/4$ period of dipole oscillations, i.e., $t_{\rm c}E_{\rm F,0}= \frac{1}{4}\frac{2\pi}{\omega_z^*}\simeq666$, where the renormalized axial trap frequency $\omega_z^*$ involves the effective mass $m_\dwn^*$ and the attractive polaron energy $\Xi_{\rm A}$ as $\omega_z^*=\omega_z\sqrt{\frac{m_\dwn}{m_\dwn^*}(1+\chi)}$.
%This estimation is valid when the spin relaxation rate is negligibly small.
%Fig.~\ref{fig2}(b) shows the critical damping behavior of impurity clouds with $\gamma_\dwn(0)=\rho=\omega_z^*$.
%One can see that the collision is largely delayed.%with increasing the temperature as $\rho\propto T^2$.
%In the overdamped regime, the impurity clouds spatially expand and slowly approach thermal equilibrium at the center of the trap.
\par
The impurity Fermi pressure $\Xi_{\rm P}$ and the polaron-polaron interaction $\Xi_{\rm F}$ induce appreciable broadening of polaronic clouds.
%The ripples also appear around the edge of impurity clouds due to these effects.
In particular, $\Xi_{\rm P}$ plays an important role just before the collision because $\Xi_{\rm P}$ becomes large when $n_\dwn$ increases.
While a core appears near the trap center ($z=0$) at $t\sim0.9t_{\rm c}$, one can find that polaronic clouds remain broadened even around $t=t_{\rm c}$.
It is analogous to the shock-wave formation observed in the collision experiment in a unitary Fermi gas~\cite{Thomas}.
\par
Figure~\ref{fig2} (b1) displays the velocity field profile $v_{z,\dwn}(z)$ with $z_{\rm I}=0.8R_z$.
Around $t=t_{\rm c}$, $|v_{z,\dwn}(z)|$ is also large in a region near the trap center where the core is seen.
The $z$-derivative of $v_{z,\dwn}(z)$ becomes discontinuous around the inner edges of the gas clouds, which is reminiscent of the shock front.
Since the shock wave is formed when $|v_{z,\dwn}(z)|$ exceeds the velocity of first sound of polaronic clouds $c_{1,\dwn}=\sqrt{\frac{n_\dwn}{m_\dwn^*}\frac{\partial^2 E}{\partial n_{\dwn}^2}}$~\cite{PinesNozieres}, which reads 
\begin{eqnarray}
c_{1,\dwn}=\sqrt{\frac{2}{m_\dwn^*}\left(\frac{\varepsilon_{\rm F,\dwn}}{3}+\zeta n_\dwn+3\tau n_\dwn^2\right)},
\label{c1down}
\end{eqnarray}
the region with a large $|v_{z,\dwn}(z)|$ is expected to be shock dominated. 
{Such a region can be identified by the Mach number profile {defined by the ratio between the local velocity field and the local speed of sound $|v_{z,\dwn}(z)|/c_{1,\dwn}(z)$} shown in Fig.~\ref{fig2} (c1).} 
%Indeed, at a typical density $n_\dwn(z)=10^{-2}n_\up(0)$, we obtain $c_{1,\dwn}\sim 2\times 10^{-4}R_zE_{\rm F,0}$.
Since $c_{1,\dwn}$ purely depends on the compressibility of polaronic clouds, 
we emphasize that this shock-wave formation exhibits a collective polaronic effect.
{The role of multi-polaron properties such as impurity Fermi pressure and induced interactions in our setup} is a sharp contrast to existing experiments relevant for single-polaron properties {such as polaron energy and effective mass}.
\par
In the case of $z_{\rm I}=0.1R_z$ shown in Figs.~\ref{fig2} {(a2), (b2), and (c2)}, on the other hand, $|v_{z,\dwn}(z)|$ does not exceed $c_{1,\dwn}$ and hence shock waves do not occur near the center. 
{The colliding velocity $v_{\rm col.}$ is crucial for the formation of {internal shock waves}.
In the case of $z_{\rm I}=0.1R_z$, a {rough estimate} gives $v_{\rm col.}\simeq\sqrt{\frac{m_\dwn}{m_\dwn}^*}\omega_z z_{\rm I} \simeq 0.11\omega_zR_z$, which is smaller than $c_{1,\dwn}\simeq \omega_zR_z\sqrt{(m_\dwn/m_\dwn^*)(Y^{\frac{2}{3}}/3+3\kappa Y/5)}\simeq0.14\omega_zR_z$ {for a small value of $Y$ considered here}. } 
Instead, shock waves appear around the outer edges of polaronic clouds just before the collision due to the growth of $|v_{z,\dwn}(z)|$ in the outer region as shown in Fig.~\ref{fig2} (b2).
Even in this case $n_{\dwn}(z)$ exhibits a maximum at $z=0$ far before $t_{\rm c}$ ($tE_{\rm F,0}\simeq200$).
This is due to the overlapping of broadened polaronic clouds.
We note that how shock waves are formed depends on not only the local sound velocity (\ref{c1down}) but also $z_{\rm I}$, $Y$, and $\eta$ in Eq.\ (\ref{eqInitial}), whereas $t_{\rm c}$ is completely determined by the single-polaron physics.
\par
If $|v_{z,\dwn}(z)|$ exceeds the velocity of first sound of the majority cloud $c_{1,\up}=\sqrt{\frac{n_\up}{m_\up}\frac{\partial^2 E}{\partial n_{\up}^2}}$, the polarons become unstable due to the Cherenkov instability~\cite{Nielsen:2019,Sekino}.
Such an ultrafast regime cannot be addressed by the present approach where the decay of a quasi-particle and the dynamics of majority clouds are neglected.
We also note that while earlier collisional studies~\cite{MIT3,Goulko:2011,Goulko:2012,Goulko:2013} treated a collision between two spins involving the direct $\up-\dwn$ attraction, our work reveals the indirect majority-mediated interaction between polarons.

%We note that in our calculations $|v_{z,\dwn}(z)|$ is always less than $\Delta z/\Delta t=3\times 10^{-2}R_zE_{\rm F,0}$, as mentioned above~\cite{CFL}. 
%In our one-dimensional set up,
%the bounce after collision is not seen due to the counter flowing current at the center of trap.
%To describe the dynamics after collision, the transverse dynamics should also be taken into account.

\begin{figure}[t]
\begin{center}
\includegraphics[width=0.9\hsize]{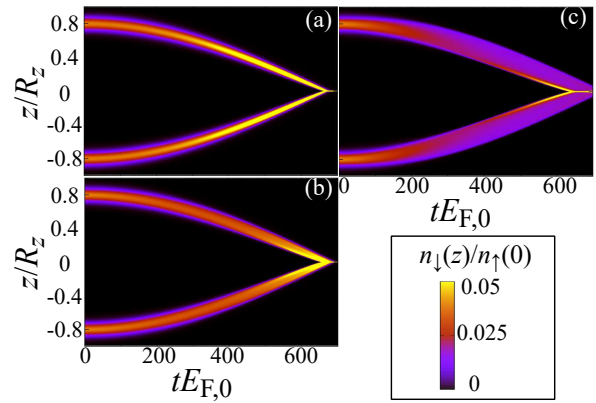}
\end{center}
\caption{Axial density profile $n_\dwn(z)$ of polaronic clouds obtained with different strength of the polaron-polaron interactions in the absence of the impurity Fermi pressure.  The cases of (a) $\kappa=0$, (b) $\kappa=0.2$, and (c) $\kappa=1$ are plotted.
}
\label{fig3}
\end{figure}

To further examine effects of $\Xi_{\rm F}$, we demonstrate the collisional dynamics of two polaronic clouds in the absence of $\Xi_{\rm P}$ {and virtually tune the induced interaction};
the results for $n_\dwn(z)$ with different strength of the polaron-polaron interactions [(a)~$\kappa=0$, (b)~0.2, and (c)~1] are exhibited in Fig.\ \ref{fig3}.
The comparison between (a) and (b) indicates that the polaron-polaron interaction with $\kappa=0.2$ %which is relevant for polarized $^6$Li-$^6$Li mixture, 
 plays a role in broadening polaronic clouds.
%although it is small compared to the impurity Fermi pressure effect shown in Fig.~\ref{fig2} (a1).
In the case of $\kappa=0.2$, the shock wave accompanied by ripples (biased peaks) can be found near the inner edge of polaronic clouds.
For $\kappa=0$, on the other hand, the shock wave is not formed during the evolution, because $\Xi_{\rm A}$ does not produce a force that involves $\nabla n_{\dwn}$ and hence acts to broaden the polaronic clouds.
In other words, two clouds in the initial setup have already behaved as shock waves because of $c_{1,\dwn}=0$ everywhere.  
In this case, two polaronic clouds propagate and shrink as they fall into the center.
To make the effect of $\Xi_{\rm F}$ more visible,
we also perform the simulation with strongly repulsive polaron-polaron interaction ($\kappa= 1$), as shown in Fig.~\ref{fig3}(c).
One can find the broadening of polaronic clouds as well as the appearance of ripples and a core.
We note that the present case without $\Xi_{\rm P}$ is relevant for bosonic impurities immersed in a Fermi sea.
%Otherwise, although we do not apparently take into account finite temperature effects in this work,
%it can be regarded as the finite temperature case with the low impurity density~\cite{TajimaUchino}
%where the impurity atoms are described by the Boltzmann statistics because of $T/T_{\rm F,\up}>1$ [$T_{\rm F,\dwn}=(6\pi^2n_{\dwn})^{\frac{2}{3}}/(2m_\dwn^*)$ is the Fermi temperature of polarons].
In the latter case, the role of $\Xi_{\rm P}$ may be replaced by the boson-boson repulsion.
%Although we neglect it for simplicity in this work, they can easily be implemented.
\par
\begin{figure}[t]
\begin{center}
\includegraphics[width=0.8\hsize]{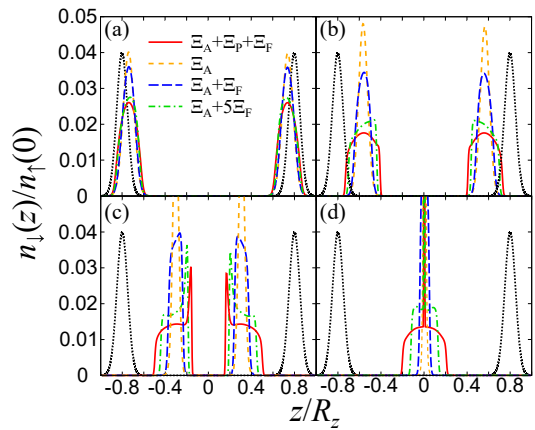}
\end{center}
\caption{Polaronic density profiles $n_\dwn(z)$ at (a) $t=t_{\rm c}/4$, (b) $t=t_{\rm c}/2$, (c) $t=(3/4)t_{\rm c}$, and (d) $t=t_{\rm c}$.
The four sets of the sum of the energy contributions are drawn as $E=\Xi_{\rm A}+\Xi_{\rm P}+\Xi_{\rm F}$, $\Xi_{\rm A}$, $\Xi_{\rm A}+\Xi_{\rm F}$, and $\Xi_{\rm A}+5\Xi_{\rm F}$.
The dotted line shows the initial density profile given by Eq.~(\ref{eqInitial}).
}
\label{fig4}
\end{figure}

Finally, we plot in Fig.~\ref{fig4} $n_{\dwn}(z)$ at $t=t_{\rm c}/4$--$t_{\rm c}$.
These results clearly reflect specific features of our protocol. 
While the comparison between the results of $E=\Xi_{\rm A}$ and $E=\Xi_{\rm A}+\Xi_{\rm F}$ at $t=t_{\rm c}/2$--$(3/4)t_{\rm c}$ shows the cloud broadening due to $\Xi_{\rm F}$, the density difference at $t=t_{\rm c}$ is negligible because the polaronic properties are important near the edge of the majority cloud rather than the trap center.
The biased peaks near the inner edge, which can be seen in the results of $E=\Xi_{\rm A}+\Xi_{\rm P}+\Xi_{\rm F}$ and $E=\Xi_{\rm A}+5\Xi_{\rm F}$ at $t=(3/4)t_{\rm c}$, clearly indicate the appearance of ripples due to the repulsive forces.
{Indeed, the density profile in Fig.~\ref{fig4}(b) is similar to {that in} the observed quantum shock wave in a Bose-Einstein condensate~\cite{Dutton}.
Since the compressive flow is characterized by the impurity Fermi pressure and induced multi-polaron interactions which do not have classical counterparts, the shock wave formation in this setup originates from quantum effects.}
From Fig.\ \ref{fig4} (d), moreover, one can confirm the shock wave collision at $t=t_{\rm c}$  in the cases of $E=\Xi_{\rm A}+\Xi_{\rm P}+\Xi_{\rm F}$ and $E=\Xi_{\rm A}+5\Xi_{\rm F}$ where the central core and surrounding parts coexist.
{Since these clouds go beyond the critical Mach number $|v_{\dwn}(z)|/c_{1,\dwn}(z)=1$ during the time evolution, the supersonic compressible flow is presumably relevant to the shock formation at a later stage.}   
These results, together with the fact that $\Xi_{\rm P}$ is well-known, suggests that one could extract information on the multi-polaron properties from relevant experiments.
\par
Here we note that while our analysis has been limited to the hydrodynamic regime, even in the unitary limit~\cite{Bruun}, the local experimental conditions can be in the collisionless regime where the relaxation time is larger than the inverse of the mode frequency~\cite{PinesNozieres}.
In such a case,
% the long-range properties of the polaron-polaron interaction would play an important role in describing the dynamics.
%In addition,
the zero sound~\cite{PinesNozieres} would be crucial instead of the first sound. 
\par
{\it Summary}--- 
In this work, we propose a new protocol to examine the collective polaron properties from collisional dynamics of polaronic clouds in cold atom experiments,  which begin by preparing the two polaronic clouds around the opposite edges of a majority gas cloud. 
%the collisional dynamics of them provides us with various polaronic properties.  
We numerically perform the hydrodynamic simulation and show how collective polaronic effects appear in the dynamics. 
In particular, we show that the collision and the shock-wave formation are sensitive to the multi-polaron correlations through the first sound of polaronic clouds{, which depends on the impurity Fermi pressure and induced multi-polaron interactions in the hydrodynamic regime due to the local thermal equilibrium.}
%the shock wave formation during the time evolution due to the impurity Fermi pressure and the repulsive polaron-polaron interaction.
In the absence of the impurity Fermi pressure and induced repulsion, polaronic clouds collapse at the trap center.
Since these effects exhibit different density dependences,
one can selectively examine each effect by tuning the experimental setup appropriately. 
\par
We emphasize that our study could open new directions for further understanding of polarons. 
The collective properties of polarons discussed in this work would give a new insight into the connection between Landau's Fermi liquid theory and recent polaron physics via the first sound.
Our scenario can immediately be applied to other systems such as Bose polarons.
In this work, we confine ourselves to the one-dimensional dynamics before the collision.
On the other hand, the quadrupole mode can be expected to occur after the collision,  especially in the region where the central core is formed.
Study of this collective mode would require full treatment of the three-dimensional dynamics.
{The dynamics of majority clouds {ignored here} would also be important for the transition from {possible} dark/bright solitons to shock waves.}
%In addition, the local density approximation adopted here neglects the long-range nature of the induced polaron-polaron interaction.
%Effects of the long-range polaron-polaron interaction could also manifest themselves in the collisonless regime with our setup, and we are now in progress along this direction~\cite{Takahashi}. 
\par
We would like to thank K. Nishimura, T. Hata, T. Hatsuda, P. Naidon, and H. Togashi for useful discussions.
H. T. is supported by a Grant-in-Aid for JSPS fellows (No.\ 17J03975).
This work is supported in part by Grants-in-Aid for Scientific Research from 
JSPS (Nos.\ 17K05445, 18K03501, 18H05406, 18H01211, and 19K14619).


\begin{thebibliography}{99}
\bibitem{note1}{
In this work, for simplicity, we use the form of Eq.~(\ref{eq1}), which is relevant in the case in which $v_\up=0$ everywhere and $v_\dwn$ is measured with respect to the majority component.
Note that the general form of Euler's equation has to possess the local Galilean invariance.
}
\end{thebibliography}
\end{document}